Ubiquitous van der Waals altermagnetism with sliding/moire ferroelectricity

Yuxuan Sheng,[1*] Junwei Liu,[2*] Jia Zhang,[1*] Menghao Wu[1,3†]

[1]School of Physics, Huazhong University of Science and Technology, Wuhan, Hubei 430074, China

[2]Department of Physics, The Hong Kong University of Science and Technology, Hong Kong, China

[3]School of Chemistry, Huazhong University of Science and Technology, Wuhan, Hubei 430074, China

*These authors contributed equally to the work.

†Email: wmh1987@hust.edu.cn

Abstract  According to the recent studies on sliding/moire ferroelectricity, most 2D van der Waals nonferroelectric monolayers can become ferroelectric via multilayer stacking. In this paper we propose that similar strategy can be used to induce desirable van der Waals altermagnetism with symmetry-compensated collinear magnetic orders and non-relativistic spin splitting. Our first-principles calculations show the pervasive co-existence of sliding ferroelectricity and altermagnetism in a series of magnetic multilayers with anti-parallel stacking configurations. Upon a twist angle in bilayers, moire ferroelectricity can be combined with altermagnetism, while some untwisted bilayers exhibit pseudo-altermagnetism with zero net magnetizations and non-relativistic spin splittings coupled with sliding ferroelectricity. Our study significantly expands the scope of altermagnetism, and its combination with sliding/moire ferroelectricity brings in new physics as well as promising applications, which should stimulate further experimental efforts.

1. Introduction

Two-dimensional (2D) materials offer a wide range of fascinating physical properties, which can also be tuned via stacking engineering of bilayers or multilayers that may even give rise to exotic effects absent in monolayers. For example, most 2D materials are non-ferroelectric, while many of them can become ferroelectric (FE) via bilayer/multilayer stacking in either parallel or antiparallel configurations, with vertical polarizations electrically switchable via in-plane sliding. Upon a twist angle, a moire superlattice of periodic FE domains can be formed.[1,2] Within a few years, such sliding/moire ferroelectricity has been experimentally confirmed in BN[3-5], transition-metal dichalcogenides (TMDs) [6-16], InSe [17] bilayers/multilayers and even organic crystals[18]. The significant technological advances due to the unique switching mode via sliding, such as ultrafast speed, low energy cost and high endurance, have also been demonstrated recently[19,20].

In this paper, we try to apply similar strategy of stacking to the design of altermagnetism, with symmetry-compensated collinear magnetic orders like antiferromagnets, and non-relativistic spin splitting like ferromagnets[21-36]. Such spin-split bands enable the control of spin channels without macroscopic magnetization, which is appealing for potential applications in spintronics owing to the advantage of non stray field and THz spin dynamics. In recent years, altermagnetism has been predicted and experimentally identified in a series of systems such as rutile fluorides/oxides $MnF_2$[37], $MnO_2$[38] and $RuO_2$[33,39,40], chalcogenide semiconductor MnTe[22,41-44], pnictides like $FeSb_2$[45], CrSb[29], etc,. It is noteworthy that a few 2D altermagnets have also been proposed[46-56], which are yet to be verified, and their scarcity can be attributed the absence of translational symmetry in the vertical direction. We also note that inversion symmetry breaking is required in both design of ferroelectricity and altermagnetism, implying the possibility of their pervasive combinations. Here through first-principles calculations we show that in some magnetic layers, altermagnetism can be formed by controlling their stacking configurations that also give rise to sliding/moire ferroelectricity. Such unprecedented van de Waals multiferroicity combining altermagnetism and sliding ferroelectricity enables nonvolatile electric-field control of magnetism within the same crystal, which is long-sought but remains elusive. [57]

2. Results and Discussion

Although most 2D monolayers possess crystal lattices with high symmetry, symmetry breaking can be induced in their bilayers via either parallel stacking antiparallel stacking. Taking transition metal dichalcogenides/dihalides $MX_2$ monolayers typically in 2H or 1T phase as the paradigmatic case, as shown in Fig. 1(a), vertical polarizations (see the black arrows) reversible via in-plane interlayer sliding can be formed in both parallel stacking of 2H layers and antiparallel stacking of 1T bilayers and multilayers, which are multiferroic if the layers are magnetic. The elementary rules for identifying altermagnetism was summarized in a previous report[30]: i) two different spin sub-lattices are connected by crystallographic rotation transformation, possibly combined with translation and inversion transformation; ii) there is no inversion center between the sites occupied by magnetic atoms; iii) the crystal present specific symmetry that belongs to specific spin symmetry group. As shown in Fig. 1(b), for 1H phase magnetic $MX_2$ monolayer, parallel-stacking layers cannot meet the first rule, while the second rule is violated in antiparallel stacking with inversion symmetry. For 1T phase, the inversion symmetry that maintains in parallel stacking will be broken in anti-parallel stacking. The symmetry breaking leads to inequivalency between two layers, so even when the two ferromagnetic (FM) layers are antiferromagnetically (AFM) coupled, the two spin sublattices are not exactly identical. However, when the vertical direction is also periodic, such interlayer inequivalency vanishes and such anti-parallel stacking bulk layers with two spin sublattices connected by 180 degree rotation transformation can comply with the first rule, and altermagnetism may emerge.

A case in point is multilayer 1T phase $NiCl_2$ that has been experimentally synthesized[58]. With the antiparallel stacking configuration shown in Fig. 2(a), the breaking of inversion symmetry gives rise to a vertical bulk polarization of 0.45 $\mu C/m^2$. In one polar state, each Ni ion is right below one Cl ion from the layer above, but not directly above one Cl ion of the layer below. Upon relative sliding of adjacent layers along the armchair direction by 1/3 lattice constant, it will be transformed to another polar state with reversed polarization where each Ni ion is right above one Cl ion of the layer below. The switching pathway between two equivalent polar states is calculated in Fig. 2(a), revealing a typical small sliding barrier of 16.4 meV/unitcell.

According to our DFT calculations, NiCl$_2$ monolayer is FM in the ground state, while for the anti-parallel stacking bulk phase in space symmetry group $P6_3mc(C_{6v})$, the interlayer couplings are AFM with spin symmetry group $P^{-1}6_3^1m^{-1}c^{\infty m}$ and magnetic space group $P'6'_3m'c$, corresponding to type SST-4B[59], and thus potentially with nonrelativistic spin splittings. The two sublattices can be connected by $A = \{6_{001}^1|\tau_{(0,0,0.5)}\}$, and the nontrivial spin space group can be obtained by group extension $G_{NSS} = \{E||G_\uparrow\} + \{U_n(\pi)||AG_\uparrow\}$, with one sublattice space group $G_\uparrow = P3m1$. The spin space group is determined by $G_{SSG} = G_{NSS} \times Z_2^k \ltimes SO(2)$[24], with $Z_2^k = \{E, TU_n(\pi)\}$ and $SO(2) = \{U_z(\phi), \phi \in [0, 2\pi]\}$. Along the $k$ path of $\Gamma - M - K - \Gamma - A - L - H - A$, the spin degeneracy is protected by symmetry as the little group of momentum k $L_k$ contains one of the symmetry transformation connecting opposite-spin sublattices (i.e. $L_k \cap \{U_n(\pi)||AG_\uparrow\} \neq \emptyset$). However, spin splitting is expectable on the surface of the triangular prism of high symmetry point in $k$ space like the path $C - L$ with 1D representation of $X_{11(1)}$, $X_{21(1)}$ for $L_k$, where $L_k \cap \{U_n(\pi)||AG_\uparrow\} = \emptyset$, consistent with our calculated band structure in Fig. 2(b). More details of the band representation are listed in Supplemental Material.

Similar combination of sliding ferroelectricity and altermagnetism may exist in various multilayer systems of antiparallel stacking 1T magnetic MX$_2$ layer, e.g., 3d metal dichachogenides like VSe$_2$ and CrSe$_2$, noting that their monolayers have been experimentally revealed to be in FM 1T phase[60,61]. In their ground states, each FM layer is AFM coupled to its adjacent layers, and the two opposing spin sublattices are connected by $A = \{6_{001}^1|\tau_{(0,0,0.5)}\}$ instead of inversion symmetry. The induced altermagnetism breaking the Kramers spin degeneracy can be revealed in their bandstructures with spin-splittings along the $k$ path of $C - L - \Gamma$ in Fig. 3(a) and (b). The inversion symmetry breaking due to the antiparallel stacking configurations give rise to sliding ferroelectricity, and the switching barriers between their bi-stable polar states are respectively 39.9 and 25.5 meV/unitcell. It is noteworthy that sliding ferroelectricity can also be formed in the antiparallel stacking hexagonal metal trihalides, e.g. H-type CrI$_3$ and CrBr$_3$ already synthesized in a previous experimental study[62], while our calculations turn out that they are FM in the ground state. In comparison, for the well-known anomalous quantum Hall effect system MnBi$_2$Te$_4$ in antiparallel stacking configurations, each FM layer is AFM coupled with its adjacent layers, where sliding ferroelectricity can also be formed with a polarization of 1.36 μC/m$^2$ and a switching barrier of 88.4 meV/unitcell. Similarly, the two opposing spin sublattices are connected by $A = \{6_{001}^1|\tau_{(0,0,0.5)}\}$, and the bandstructure in Fig. 3(c) exhibits spin-splittings along the $k$ path of $C - L - \Gamma$.

As mentioned above, due to the interlayer inequivalency of sliding ferroelectric bilayers, the two spin sublattices are not exactly equivalent even when the two layers are antiferromagnetically coupled. However, the two layers will become identical upon a twist angle that leads to zero net polarizations, and the two spin sublattices can be connected by rotation operation for a magnetic bilayer with antiferromagnetic interlayer coupling. Such stacking strategy is not only applicable to 1T monolayers, but also 2H monolayers. For example, such antiferromagnetism may exist in twisted 2H VS$_2$ bilayer with spin space group (SSG) $P^13^{-1}2^11^{\infty m}1$, and two opposing spin sublattices are connected by $A = \{-1||2_{010}\}$. As shown in the bandstructure of Fig 4(a), spin splitting is absent along the high symmetry point $\Gamma - M - K - \Gamma$ path, which emerges in the path of $M/2\left(\frac{1}{4}, 0, 0\right) - K/2\left(\frac{1}{6}, \frac{1}{6}, 0\right)$. Such altermagnetism can be combined with moire ferroelectricity as a moire superlattice of ferroelectric domains can be formed upon a small twist angle in a sliding ferroelectric bilayer (see Fig. 4(b)). For untwisted bilayers, despite the inequivalency induced by vertical polarizations leading to spin-splittings, many magnetic sliding ferroelectric bilayers possess zero net magnetizations, e.g., parallel-stacking 2H VS$_2$ bilayer in Fig. 4(c). Meanwhile its vertical polarization (~0.71 pC/m) elevates the bands of one layer away from the other (by around 50 meV at Gamma point), breaking the spin degeneracy even the net magnetization is zero. With compensated collinear magnetic orders and non-relativistic spin splitting, such spin configuration might be denoted as pseudo-altermagnetism that actually possesses most merits of typical altermagnetism. Moreover, it is coupled with sliding ferroelectricity as the spin-splitting $E_{\text{spin-up}}$-$E_{\text{spin-down}}$ can be reversed upon ferroelectric switching via interlayer sliding, which can be verified in the bandstructures in Fig. 4(c).

## 3. Conclusions

In summary, we show first-principles design of ubiquitous van der Waals altermagnetism combined with sliding/moire ferroelectricity within the same crystal. Such unique multiferroicity does not only significantly broaden the scope of both research fields, but also render promising applications combining the advantages of both properties, noting the recent breakthroughs on sliding ferroelectricity that reveal the ultrafast speed and high endurance comparable to the state-of-the-art ferroelectrics[19,20].

## 4. Methods

We performed first-principles calculations based on density functional theory (DFT) methods implemented in the Vienna Ab initio Simulation Package (VASP 5.4.4) code[63,64]. The exchange–correlation effect described within the generalized gradient approximation in the Perdew–Burke–Ernzerh[65]functional, together with the projector augmented wave[66]method, are adopted. The kinetic energy cutoff is set to be 600eV, and the GGA+$U_{effective}$[64] approach is introduced to deal with localized d orbitals, with U=3.0 eV for $VS_2$, U=4.47eV for $NiCl_2$[67], U=4.0 eV and J=0.9 eV for $MnBi_2Te_4$[68], U=4.5 eV and J=0.6 eV for $CrSe_2$[69], following previous studies. The van der Waals correction of Grimme with Becke-Johnson damping function [70] is applied, except for $CrSe_2$ systems where correction of optB86b functional for the exchange potential[71] (optB86b-vdW) is used following a previous study [69]. For bilayer systems, a vacuum space of 30 Å is set to avoid the artificial interaction induced by period boundary conditions. All the structures are fully relaxed until the energy and forces converged to $10^{-6}$ eV and $0.01$ eV·Å$^{-1}$, respectively. The solid state nudged-elastic-band[72] (SS-NEB) method and Berry phase method[73] are respectively used for calculating the ferroelectric switching pathways and polarizations.

Data available statements

The data in this study is available upon reasonable request.

Acknowledgements

This work is supported by National Natural Science Foundation of China (Nos. 22073034).

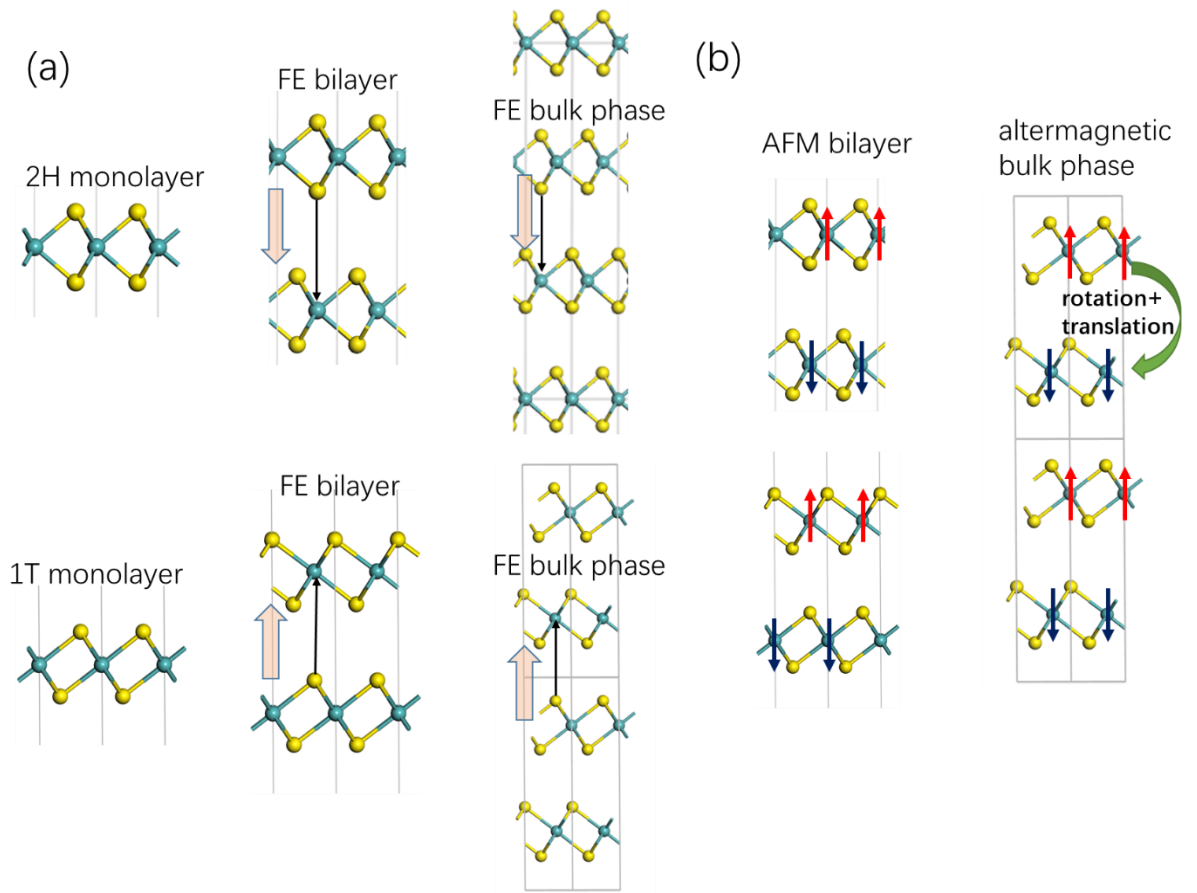

Figure 1. (a) Geometric structures of 2H/1T monolayers, their sliding ferroelectric bilayers and bulk phases. (b) Among their corresponding AFM states, the bulk 1T phases can potentially become altermagnetic with two spin sublattices connected by rotation+translation. The red and blue arrows respectively denote spin-up and spin-down, while orange arrows denote electrical polarization directions.

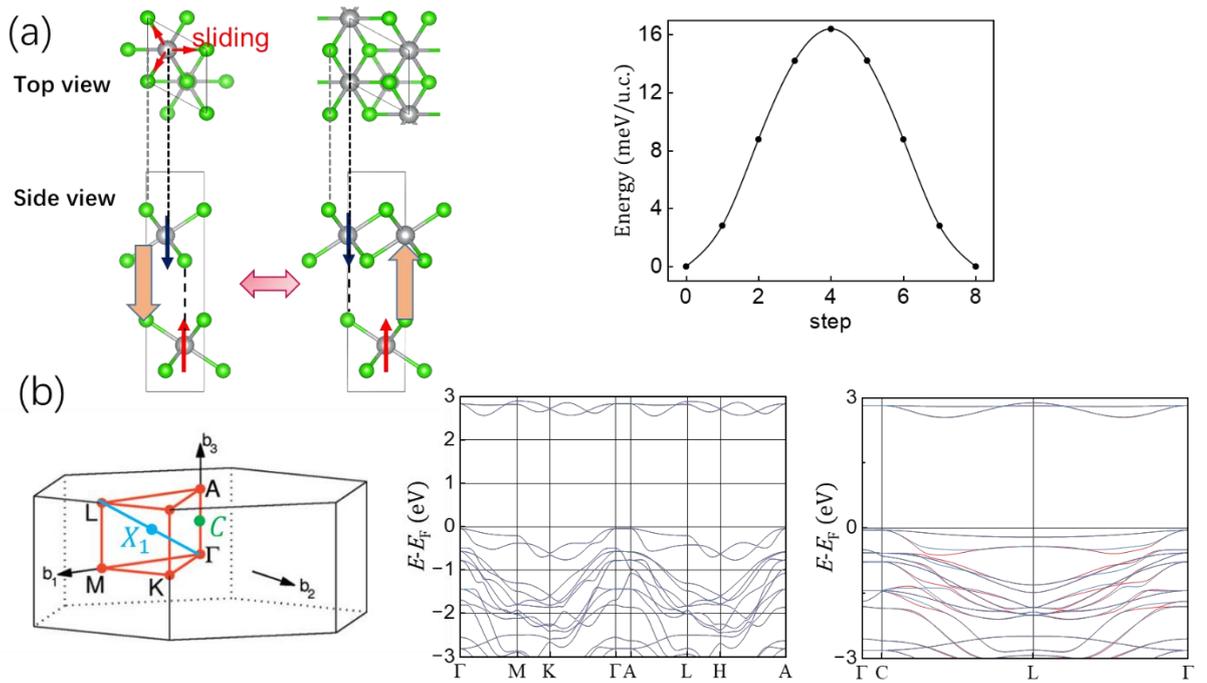

Figure 2. (a) Geometric structures of bulk NiCl$_2$ in antiparallel stacking configurations and the switching pathway between two polar bistable states. (b) The corresponding bandstructure along different k paths, where the red and black bands respectively denote spin-up and down channels.

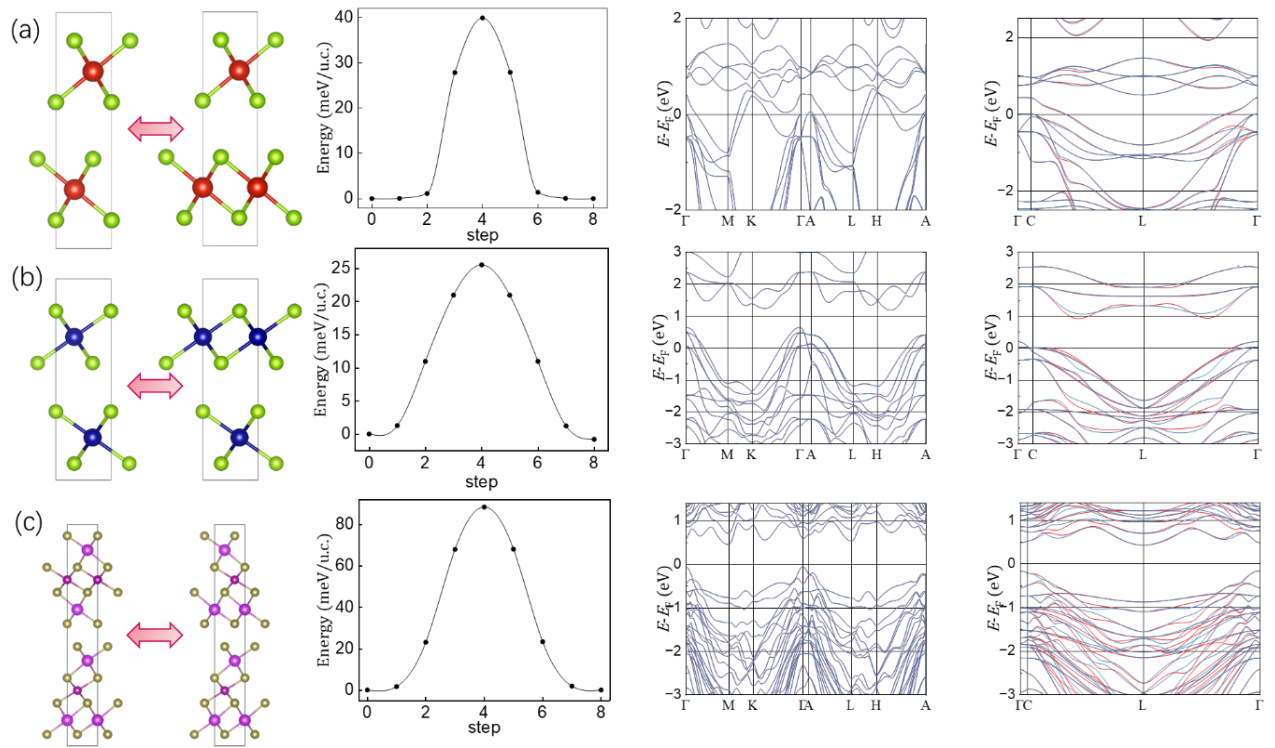

Figure 3. Ferroelectric switching pathways and bandstructures for bulk (a) VSe$_2$, (b) CrSe$_2$, and (c) MnBi$_2$Te$_4$ in antiparallel stacking configurations.

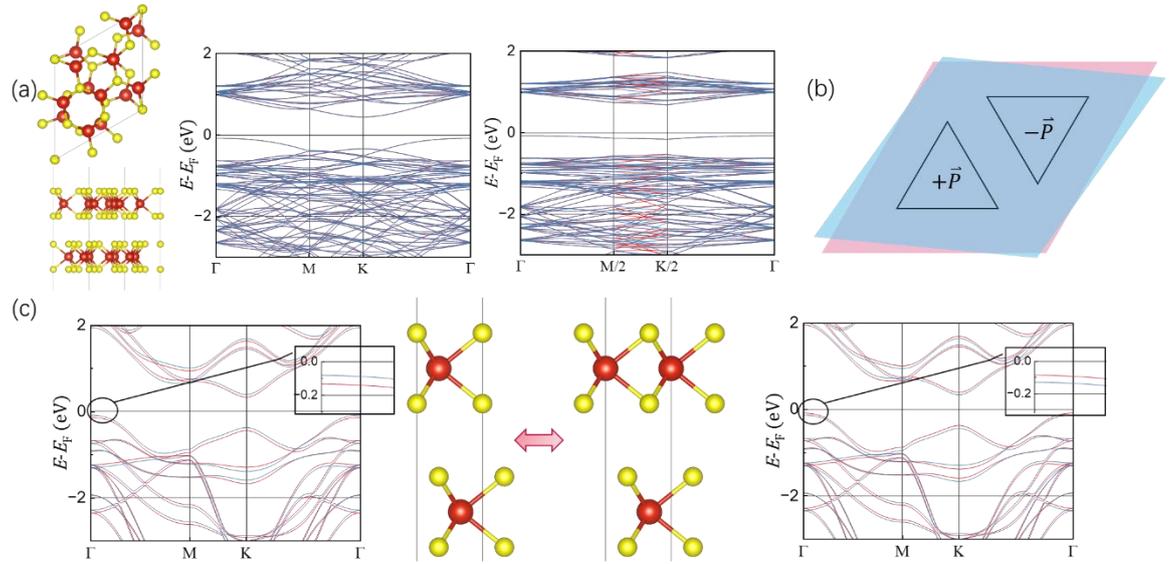

Figure 4. (a) Geometric structure and bandstructure of twisted bilayer 2H $VS_2$. (b) An illustration of moire ferroelectric domains upon a small twist angle. (c) The change of spin-splittings in bandstructure upon ferroelectric switching for non-twisted bilayer 2H $VS_2$.